\documentclass[nofootinbib,prd,aps,groupedaddress,longbibliography
]{revtex4-1}

\usepackage{bbm}
\usepackage{color}
\usepackage{graphicx}
\usepackage{amssymb,latexsym}
\usepackage{tensor}
\usepackage{slantsc}
\usepackage{slashed}

\usepackage{mathtools}
\usepackage[normalem]{ulem}

\usepackage{marginnote}

\newcommand{\Dnabla}{D}
\newcommand{\tWang}{{W}}
\newcommand{\redR}{R}
\newcommand{\sigmaH}{\sigma_H}
\newcommand{\redRH}{R_H}
\newcommand{\red}[1]{{\color{red} #1}}
\newcommand{\nored}[1]{{#1}}

\newtheorem{theorem}{\sc  Theorem\rm}[section]

\newtheorem{Theorem}[theorem]{\sc  Theorem\rm}

\newtheorem{Lemma}[theorem]{\sc Lemma\rm}

\newtheorem{remark}[theorem]{\sc Remark\rm}
\newtheorem{Remark}[theorem]{\sc Remark\rm}

\DeclareFontFamily{OT1}{rsfs}{}
\DeclareFontShape{OT1}{rsfs}{m}{n}{ <-7> rsfs5 <7-10> rsfs7 <10-> rsfs10}{}
\DeclareMathAlphabet{\mycal}{OT1}{rsfs}{m}{n}

{\catcode `\@=11 \global\let\AddToReset=\@addtoreset}
\AddToReset{equation}{section}

\newcounter{mnotecount}[section]

\newcommand{\T}{\mathbb{T}}

\newcommand{\eel}[1]{\label{#1}\end{equation}}
\newcommand{\eeal}[1]{\label{#1}\end{eqnarray}}

\newcommand{\bel}[1]{\begin{equation}\label{#1}}
\newcommand{\bea}{\begin{eqnarray}}
\newcommand{\bean}{\begin{eqnarray}\nonumber}
\newcommand{\beal}[1]{\begin{eqnarray}\label{#1}}
\newcommand{\eea}{\end{eqnarray}}

\newcommand{\nn}{\nonumber}

\newcommand{\qed}{\hfill $\Box$}

\newcommand{\proof}{\noindent {\sc Proof:\ }}
\newcommand{\be}{\begin{equation}}
\newcommand{\eeq}{\end{equation}}
\newcommand{\ee}{\end{equation}}
\newcommand{\beqa}{\begin{eqnarray}}
\newcommand{\eeqa}{\end{eqnarray}}
\newcommand{\beqan}{\begin{eqnarray*}}
\newcommand{\eeqan}{\end{eqnarray*}}
\newcommand{\ba}{\begin{array}}
\newcommand{\ea}{\end{array}}

\newcommand{\mnote}[1]
{\protect{\stepcounter{mnotecount}}$^{\mbox{\footnotesize
$
\bullet$\themnotecount}}$ \marginpar{
\raggedright\tiny\em
$\!\!\!\!\!\!\,\bullet$\themnotecount: #1} }

\newcommand{\warn}[1]
{\protect{\stepcounter{mnotecount}}$^{\mbox{\footnotesize
$
\bullet$\themnotecount}}$ \marginpar{
\raggedright\tiny\em
$\!\!\!\!\!\!\,\bullet$\themnotecount: {\bf Warning:} #1} }

\newcommand{\R}{\mathbb R}

\newcommand{\beaa}{\begin{eqnarray*}}
\newcommand{\eeaa}{\end{eqnarray*}}

\def\ben{\begin{equation}}
\def\een{\end{equation}}
\def\bena{\begin{eqnarray}}
\def\eena{\end{eqnarray}}

\def\f(#1/#2){\frac{#1}{#2}}
\def\\tWangrac(#1/#2){\left(\frac{#1}{#2}\right)}
\def\chris(#1-#2-#3){{\mit \Gamma}^{#1}{}_{{#2}{#3}} }
\def\tilchris(#1-#2-#3){\tilde{{\mit \Gamma}}^{#1}{}_{{#2}{#3}}}
\def\hatchris(#1-#2-#3){\hat{{\mit \Gamma}}^{#1}{}_{{#2}{#3}}}

\begin{document}

\title{Uniqueness and energy bounds for static AdS metrics\footnote{Preprint UWThPh-2019-25}}

\author{Piotr T.\ Chru\'{s}ciel}
\email[]{piotr.chrusciel@univie.ac.at}
\homepage[]{http://homepage.univie.ac.at/piotr.chrusciel}
\thanks{supported in part by the Polish National Center of Science (NCN) 2016/21/B/ST1/00940 and the Austrian Science Foundation, FWF project P 29517-N27.}
\affiliation{Faculty of Physics, University of Vienna}

\author{Gregory J. Galloway}
\email[]{galloway@math.miami.edu}
\thanks{supported in part by NSF grant DMS-1710808.}
\affiliation{University of Miami}

\author{Yohan Potaux}
\email[]{yohan.potaux@ens-lyon.fr}
\affiliation{ENS de Lyon}

\date{\today}

\begin{abstract}
We show that Wang's proof of uniqueness of Anti-de Sitter spacetime can be adapted to provide   uniqueness results for strictly static asymptotically locally hyperbolic vacuum metrics with toroidal infinity, and to prove negativity of the  free energy  $E-TS$ of asymptotically AdS black holes with higher-genus horizons.
\end{abstract}

\pacs{}

\maketitle

\tableofcontents

\section{Introduction}

In~\cite{wang_uniqueness_2002} Wang derived an identity which allowed him to prove, in all space-dimensions $n\ge 3$, uniqueness of anti-de Sitter space within the class of strictly static conformally compactifiable solutions of the vacuum Einstein equations with a negative cosmological constant and a spherical conformal infinity. (The identity already appears in \cite{Shen}.) The aim of this note is to explore further consequences of this identity. Namely, we consider static solutions of the vacuum Einstein equations with a  negative cosmological constant and prove:

\begin{enumerate}
  \item The cuspidal
   Birmingham-Kottler metrics are unique in the class of strictly static solutions containing ALH ends and other controlled asymptotic ends
      or suitable boundaries, cf.\ Theorems~\ref{T11VI19.3}-\ref{T25VI19.1}  and \ref{rigid}
      below.
  \item We establish a new lower bound for entropy of horizons in terms of the genus of the horizon, cf.\ Equation~\eqref{AreaBoundNeg} below.
  \item We prove an upper bound, suggested in~\cite{CGW}, on the free energy of a connected static black hole in terms of the genus of the horizon, cf.\ Equation~\eqref{7VII19.4} below.
\end{enumerate}

When $\Lambda$ is positive we
review the argument of \cite{Shen} (compare \cite{ambrozio_static_2015}),  that the identity mentioned above provides a simple proof of an upper bound for entropy of horizons, Equation~\eqref{AreaBoundPos} below. This upper bound has been rediscovered
in \cite{boucher_uniqueness_1984,BorghiniMazzieri1}, with different proofs.

This work can be thought-of as a continuation of~\cite{GallowayWoolgar}.  It reviews, in a slightly different manner, some results from~\cite{GallowayWoolgar} and considers further applications.
Similar ideas have been independently pursued in~\cite{LanHyun}.

\section{The equations}

We consider the vacuum Einstein equations with a cosmological constant $\Lambda$ for a static spacetime metric which we denote by $\bar g$:
\begin{equation}\label{22VII19.9}
  \bar g = -V^2 dt^2 + \underbrace{g_{ij} dx^i dx^j}_{=:g}
  \,,
  \qquad
  \partial_t V = \partial_t g = 0
  \,.
\end{equation}
We always assume that $V\not \equiv 0$
and that $(M,g)$ is complete, possibly with boundary.
We shall say that $(M,g,V)$ is strictly static if $V>0$ on $M$.

It is convenient to rescale the metric by a constant to obtain
\begin{equation}
 \Lambda = \varepsilon \frac{n(n-1)}{2}
 \,,
\label{lambdanorm}
\end{equation}
with $\varepsilon  \in \{-1,0,1\}$ according to the sign of $\Lambda$
 \footnote{Wang assumes $\Lambda < 0$ so he always has $\varepsilon = -1$.}. This leads to the following equations, where $\Dnabla $ is the covariant derivative of $g$,  $\bar R_{\alpha\beta}$ is the Ricci tensor of $\bar g$ and $R_{ij}$ the Ricci tensor of $g$:
\begin{eqnarray}
 &
 \bar R_{\alpha\beta} = \varepsilon  n\bar{g}_{\alpha\beta}
 \,,
 &
\label{Einstein2norm}
\\
 &
 \Dnabla _i\Dnabla _jV = V\big(R _{ij} - \varepsilon  ng_{ij}\big)
 \,,
 &
\label{Einsteingnorm}
\\
 &
 \Delta V = - \varepsilon  nV
 \,,
 &
\label{EinsteinVnorm}
\\
 &
 R \equiv g^{ij} R_{ij}= \varepsilon  n(n-1)
 \,.
 &
\label{Rgnorm}
\end{eqnarray}
By an abuse of terminology, triples $(M,g,V)$ satisfying the above will also be called solutions of the static vacuum Einstein equations.

We note that there cannot be a static solution $(M,g,V)$ with $\epsilon=-1$ on a compact manifold, otherwise \eqref{EinsteinVnorm} and the maximum principle imply   $V\equiv 0$, contradicting the definition of staticity.

\section{The divergence identity}
 \label{s23VII19}

 As in~\cite{GallowayWoolgar}, the key to our analysis is an identity which has been used by Shen~\cite{Shen} in dimension three in a  context related to ours, and  by Wang~\cite{wang_uniqueness_2002} in all dimensions $n\ge 3$ to  prove uniqueness of anti-de Sitter spacetime.  For the convenience of the reader we rederive this identity here.

%

We define the symmetric tensor field $\tWang$ on $M$ by%
\footnote{In his paper Wang denotes his tensor by $T$ but we use $\tWang$ in order to avoid confusion with the stress-energy tensor.}
\begin{equation}
 \tWang_{ij} \equiv \redR _{ij} - \varepsilon  (n - 1)g_{ij} = {V}^{-1}
 \Dnabla _i\Dnabla _jV + \varepsilon  g_{ij}
 \,,
\end{equation}
where the last equality, which follows from \eqref{Einsteingnorm}, holds on the set where $V$ does not vanish. Now, we have
\begin{eqnarray}
 V|\tWang|_g^2 & = & V\tWang_{ij}g^{ik}g^{jl}\tWang_{kl}
 \nn
\\
 & = & \tWang_{ij}\Dnabla ^i\Dnabla ^jV - \varepsilon \tWang_{ij}g^{ij}
 \nn
\\
 & = & \tWang_{ij}\Dnabla ^i\Dnabla ^jV -
 \varepsilon  \big(\redR  - \varepsilon  n(n-1)\big)
 \nn
\\
 & = & \tWang_{ij}\Dnabla ^i\Dnabla ^jV
 \,,
 \label{6IX19.99}
\end{eqnarray}
where in the last step we used \eqref{Rgnorm} and where $|\tWang|_g^2 = \langle \tWang, \tWang \rangle_g$ is the squared norm of $\tWang$.
Thus
\begin{eqnarray}
 V|\tWang|_g^2 & = &  \tWang_{ij}\Dnabla ^i\Dnabla ^jV
 \,.
 \label{6IX19.100}
\end{eqnarray}
Note that the calculation \eqref{6IX19.99} is valid only on the region where $V$ has no zeros. But in
\eqref{6IX19.100} both sides are smooth everywhere. Furthermore, it is well known that the set where $V$ does not vanish is dense. This implies that \eqref{6IX19.100} is true throughout $M$, regardless of zeros or sign of $V$.
 Since $(M,g)$ is Riemannian, $|\tWang|_g^2$ is {nonnegative}  and equal to zero if and only if $\tWang=0$. The last identity implies
\begin{eqnarray}
 V|\tWang|_g^2 & = & \tWang_{ij}\Dnabla ^i\Dnabla ^jV
 \nn
\\
 & = & \Dnabla ^i(\tWang_{ij}\Dnabla ^jV) - (\Dnabla ^i\tWang_{ij})(\Dnabla ^jV)
 \nn
\\
 & = &\Dnabla ^i(\tWang_{ij}\Dnabla ^jV)
 \,,
  \label{31VIII19.1}
\end{eqnarray}
since
\begin{eqnarray}
 \Dnabla ^i\tWang_{ij} & = &
 \Dnabla ^i \redR _{ij} - \varepsilon  (n-1)\Dnabla ^i g_{ij}
 \nn
\\
 & = & \Dnabla ^i \redR _{ij}
 \nn
\\
 & = & \frac{1}{2}\Dnabla _j \redR
 \nn
\\
 & = & 0
 \,,
\end{eqnarray}
where we  used  again \eqref{Rgnorm}. We can integrate over $M$ with the measure d$\mu_g = \sqrt{\mathrm{det}(g)}$ to get
\begin{eqnarray}
 \int_MV|\tWang|_g^2\,\mathrm{d}\mu_g & = &
 \int_M\Dnabla ^i(\tWang_{ij}\Dnabla ^jV)\,\mathrm{d}\mu_g
 \nn
\\
 & = & \int_{\partial M}\tWang_{ij}\Dnabla ^iVN^j\,\mathrm{d}\sigma
 \,,
\end{eqnarray}
where we applied Stokes' theorem with $\partial M$ the boundary of $M$, $d\sigma$ the measure on $\partial M$ and $N$ the unit outer directed normal vector field of $\partial M$.

Now, let us suppose  that $V$ is positive on the interior of $M$ and that $V$ vanishes on its boundary
$$
 H \equiv \left\{p \in M ; V(p)=0\right\}
  \,,
$$
with $H$ not necessarily connected.
We further assume that  $M$ has a  conformal boundary at infinity $ \partial M_\infty$  (which we always assume to be compact, but not necessarily connected), and write
\begin{equation}
 \partial M = \partial M_\infty \cup H
 \,.
 \label{boundaryM}
\end{equation}
 We have
\begin{eqnarray}
 \int_MV|\tWang|_g^2\,\mathrm{d}\mu_g & = &
\int_{\partial M_\infty}\tWang_{ij}\Dnabla ^iVN^j\,\mathrm{d}\sigma
 \nn
\\
 & + & \int_{H}\tWang_{ij}\Dnabla ^iVN^j\,\mathrm{d}\sigmaH
\,,
\label{integration}
\end{eqnarray}
where we assumed that $M\cup \partial M_\infty$ is compact, and where the integral over the conformal boundary at infinity is understood by a limiting process.
We denote by $d\sigmaH $ the measure induced by $g$ on $H$, and continue to denote by $d\sigma$ the limiting measure arising on the boundary at infinity of $M$ in the limit.

\section{The integral over the horizon}

The integral over the horizon $H$ in \eqref{integration} has been rewritten in a convenient form in~\cite{GallowayWoolgar}. We rederive the formula for completeness.
 Recall  that the surface gravity $\kappa = \sqrt{g(\Dnabla  V, \Dnabla  V)}|_H$ of $H$ is constant on each connected component  $H_p$
of $H=\cup_{p=1}^P H_p$, for some $P\in \mathbb N$. Thus there exists a locally constant function $\kappa : H \rightarrow \mathbb{R}^{+*}:= \mathbb{R}^{+} \setminus \{0\} $
 such that on $H$
\begin{equation}
 |\Dnabla  V|_g = \kappa
 \,.
\end{equation}
Then, denoting by $N$ the outer normal to $H$, on each connected component $H_p$ we have
\begin{equation}
 N = - \frac{\Dnabla  V}{|\Dnabla  V|_g} = -\frac{\Dnabla  V}{\kappa_p}
 \,,
\end{equation}
where $\kappa_p \in \mathbb{R}^{+*}$ is the value of $\kappa$ on $H_p$ and the minus sign comes from the fact that $V$ decreases approaching $H$ as $V \geq 0$ on $M$ and $V=0$ on $H$. Thus
\begin{eqnarray}
 \lefteqn{\int_{H}\tWang_{ij}\Dnabla ^iVN^j\,\mathrm{d}\sigmaH }
 \nn
\\
 & = & -\sum_{H_p}\frac{1}{\kappa_p}\int_{H_p}\tWang_{ij}\Dnabla ^iV\Dnabla ^jV\,\mathrm{d}\sigmaH
 \nn
\\
 & = & -\sum_{H_p}\frac{1}{\kappa_p}\int_{H_p}(\redR _{ij} - \varepsilon  (n-1)g_{ij})\Dnabla ^iV\Dnabla ^jV\,\mathrm{d}\sigmaH
 \nn
\\
 & = & -\sum_{H_p}\frac{1}{\kappa_p}\int_{H_p}(\redR _{ij}\Dnabla ^iV\Dnabla ^jV - \varepsilon  (n-1)\kappa_p^2)\,\mathrm{d}\sigmaH
 \,.
 \nn
 \\
 \label{integwithricci}
\end{eqnarray}

We denote by $\red{g_H}$ the metric induced by $g$ on $H$.
Letting $\redRH $ denote the Ricci scalar of the metric ${g_H}$, we will need the Gauss embedding equation
\begin{equation}
 \redRH  = \redR  - 2g(N,N)\redR _{ij}N^iN^j + g(N,N)((h^{ij}A _{ij})^2 - |A |_h^2)
 \label{scalarconstraint}
 \,,
\end{equation}
where   $A $ is the extrinsic curvature tensor of $H$ in $M$, defined for two vector fields $X,Y$ tangent to $H$ as
\begin{equation}
 A (X,Y) = g(\Dnabla _XN,Y)
 \,.
\end{equation}
It is well known that $H$ is totally geodesic, i.e. $A\equiv 0$, which can  be seen as follows:
\begin{eqnarray}
 A (X,Y) & = &-\frac{1}{\kappa}g(\Dnabla _X \Dnabla  V,Y)
 \nn
\\
 & = & -\frac{1}{\kappa}g_{ij}X^k\Dnabla _k\Dnabla ^iVY^j
 \nn
\\
 & = & -\frac{1}{\kappa}X^kY^j\Dnabla _k\Dnabla _jV
 \nn
\\
\nonumber
 & = & -\frac{1}{\kappa}X^kY^j V\left(\redR _{kj} - \frac{2\Lambda}{n-1}g_{kj}\right)
\\
 & = & 0
 \,,
 \label{extrinsiciszero}
\end{eqnarray}
since $V$ is zero on $H$. Thus $A =0$ and \eqref{scalarconstraint} becomes, using $g(N,N)=1$ and $N =- {\Dnabla  V}/{\kappa}$,
\begin{equation}
 \redRH  = \redR  - \frac{2}{\kappa^2}\redR _{ij}\Dnabla ^iV\Dnabla ^jV
 \,.
\end{equation}
We can now rewrite \eqref{integwithricci} as
\begin{eqnarray}
 \nonumber
\lefteqn{
 \int_{H}\tWang_{ij}\Dnabla ^iVN^j\,\mathrm{d}\sigmaH
}
&&
\\
&
 = & -\sum_{H_p}\frac{1}{\kappa_p}\int_{H_p}\left(\frac{\kappa_p^2}{2}(\redR -\redRH ) - \varepsilon  (n-1)\kappa_p^2\right)\,\mathrm{d}\sigmaH
 \nn
\\
 & = & \sum_{H_p}\frac{\kappa_p}{2}\int_{H_p}(\redRH  - \redR  + 2\varepsilon (n-1))\,\mathrm{d}\sigmaH
 \nn
\\
 & = & \sum_{H_p}\frac{\kappa_p}{2}\int_{H_p}(\redRH  -\varepsilon  n(n-1) + 2\varepsilon (n-1))\,\mathrm{d}\sigmaH
 \nn
\\
 & = & \sum_{H_p}\frac{\kappa_p}{2}\int_{H_p}(\redRH  - \varepsilon (n-1)(n-2))\,\mathrm{d}\sigmaH
 \,.
\end{eqnarray}
Using this result we obtain the key identity
\begin{eqnarray}
 \lefteqn{\int_MV|\tWang|_g^2\,\mathrm{d}\mu_g}
 \nn
\\
 & = & \int_{\partial M_\infty}\tWang_{ij}\Dnabla ^iVN^j\,\mathrm{d}\sigma
 \nn
\\
 & + & \sum_{H_p}\frac{\kappa_p}{2}\int_{H_p}(\redRH  - \varepsilon (n-1)(n-2))\,\mathrm{d}\sigmaH
 \,.
\label{integration2}
\end{eqnarray}
%

\section{The boundary term at infinity}

To avoid ambiguities, we emphasise that in this section $\varepsilon = -1$.

A Hamiltonian analysis of general relativity leads,  after many integrations by parts, to the following
formula for the mass of an  asymptotically locally hyperbolic end \cite{BCHKK}
 \footnote{We note a misprint in \cite[Equation~(4.40)]{BCHKK}, where a prefactor $1/16$ should be replaced by $1/8$.}
(compare~\cite{HerzlichRicciMass})
\begin{equation}
  m
    =
   - \frac{1}{8 (n-2)\pi}
   \lim_{r\rightarrow\infty}\int_{r=\mbox{\scriptsize const}}  \nabla ^j V
    ( R^i{}_j - \frac R n \delta^i_j)
    d\sigma_i
     \,,
           \label{8VII19.1}
\end{equation}
where the multiplicative prefactor in front of the integral arises from the Hilbert Lagrangean $\bar R/(16 \pi)$, as relevant for the physical spacetime dimension $n+1=4$.

%
It follows that the integral over the conformal boundary at infinity in \eqref{integration2} is related to the  total  mass $m$  (i.e., the sum of the masses over all ALH ends)  of the spacetime as
%
\begin{equation}
 \int_{\partial M_\infty}\tWang_{ij}\Dnabla ^iVN^j\,\mathrm{d}\sigma =
 \nored{{-8(n-2) m \pi }}
 \,,
\label{Wang}
\end{equation}
and thus we get
\begin{eqnarray}
 \lefteqn{\int_MV|\tWang|_g^2\,\mathrm{d}\mu_g}
 \nn
\\
 & = &   \nored{{-8(n-2) m \pi }}
 \nn
\\
 & + & \sum_{H_p}\frac{\kappa_p}{2}\int_{H_p}(\redRH  +(n-2)(n-1))\,\mathrm{d}\sigmaH
 \,.
\end{eqnarray}

Recall, an ALH static triple $(M,g,V)$ is \emph{strictly static} if $V$ is positive on $M$.
In the strictly static,
 conformally compact
 and boundaryless case we obtain
\begin{equation}
 \int_MV|\tWang|_g^2\,\mathrm{d}\mu_g
 = - \nored{8(n-2) m \pi }
 \,.
\end{equation}
Since the left-hand side is  {nonnegative}, we
recover a result of \cite{GallowayWoolgar}:

\begin{Theorem}
 \label{T11VI19.1}
Consider a  strictly static solution of the static Einstein equations $(M,g,V)$ with negative cosmological constant on a conformally compact  manifold without boundary. Then the total mass is negative or zero, vanishing if and only if $(M,g)$ is the hyperbolic space.
\end{Theorem}

The fact that the vanishing of the mass implies hyperbolic space is justified as follows: When the mass vanishes, the divergence identity shows that the metric is Einstein. Thus the Hessian of $V$ is proportional to the metric, which is the well studied Obata's equation.  It follows e.g.\ from \cite[Theorem~2]{Tashiro} that all complete metrics for which $DV$ has no zeros are \emph{not} compactifiable (compare \cite[Proposition~4.2]{GallowayWarpedSplit}). We conclude that, under the current assumptions,  $DV$ must have a zero, which leads to hyperbolic space again by  \cite{Tashiro}.

As emphasized in \cite{wang_uniqueness_2002}, Theorem~\ref{T11VI19.1} leads to uniqueness of the anti-de Sitter spacetime, which has spherical conformal infinity, and thus non-negative mass by \cite{chrusciel_hyperbolic_2019}. (Wang refers to \cite{Wang, AnderssonGallowayCai,ChHerzlich} for positivity results; these last papers contain restrictive hypotheses, which have been meanwhile removed through the work in \cite{Lohkamp2,chrusciel_hyperbolic_2019,SchoenYau2017}.
See also \cite{HuangJangMartin} for the rigidity case of these positive mass theorems, where spherical conformal infinity is assumed.)
Note that examples of metrics, as in the theorem, with negative mass are provided by the Horowitz-Myers metrics.
The theorem shows that if any further such solutions exist, they would have to have non-spherical infinity and negative mass.

Related negativity results for the mass, in the spirit of Theorem~\ref{T11VI19.1}, can be found in~\cite{ChruscielSimon,GallowayWoolgar}.

\section{Uniqueness theorems for the cuspidal Birmingham-Kottler metrics}
 \label{s3VII19.1}

In this section we continue to assume that  $\varepsilon = -1$.

Both the technique of the proof and the argument generalize to cover somewhat more general geometries, which we describe now.
The \emph{cuspidal Birmingham-Kottler (BK) metrics}  provide  a guiding example. The metric can be written in the form
%
\begin{equation}\label{2VII19.21}
  g = \frac{dr^2}{V^2} + r^2 h
   \,,
   \qquad
   V = r
   \,,
\end{equation}
where $h$ is a Ricci-flat metric on a compact $(n-1)$-dimensional manifold $B$. One checks that $g$ is Einstein,
and has zero mass in the \emph{asymptotically locally hyperbolic (ALH) end}, defined as the region where $r$ tends to infinity. The underlying manifold $\R\times B$  has two asymptotic regions, with the already mentioned ALH end and a \emph{complete cuspidal end} in which $r$ tends to zero.

The example suggests a  natural generalisation of  Wang's argument to manifolds which contain two kinds of asymptotic regions: the usual asymptotically locally hyperbolic ones, as well as ends with mildly controlled asymptotic behaviour, as captured by the following definition:
 We will say that a triple $(M,g,V)$ is  \emph{asymptotically locally hyperbolic with mild ends} if $(M,g)$ is a complete manifold which admits  an exhaustion $M=\cup_{i\in \mathbb N}M_i$  by smooth compact manifolds $M_{i}\subset M_{i+1}$ with boundaries
$$
 \partial M_i = \partial_1 M_i \cup \partial_2 M_i
$$
where the (not necessarily connected) boundaries $\partial_2 M_i$ are a union of smooth hypersurfaces which approach a (compact) conformal boundary at infinity of $M$, while the (not necessarily connected) boundaries $\partial_1 M_i$ are a union of smooth hypersurfaces  on which
\begin{equation}\label{2VII19.22}
  |dV|_g\big|_{\partial_1 M_i} \times  |\tWang|_g\big|_{\partial_1 M_i} \times A(\partial_1 M_i) \to_{i\to \infty} 0
  \,,
\end{equation}
%
where $A(\partial_1 M_i)$ is the ``area'' of the submanifolds $\partial_1 M_i$. Here $| \cdot |_g$ denotes the norm with respect to the metric $g$,
and we assume that the number of boundary components is bounded by a number independent of $i$.
The ``mild ends'' are then the regions associated with the boundaries satisfying \eqref{2VII19.22}.

As formulated so far, the definition allows some  ALH ends to be mild ends. This occurs for example for hyperbolic space, where $W\equiv 0$. To avoid this issue, which would lead to the need to add annoying trivial comments when formal statements are made, we add to the definition of a mild end the requirement that a mild end is \emph{not} ALH.

The conditions above
are clearly satisfied by the metric \eqref{2VII19.21}, where both $A(\partial_1 M_i)$ and $|dV|_g\big|_{\partial_1 M_i}$ tend to zero when $\partial_1 M_i$ is taken to be $\{r=1/i\}$, $1\le i\in \mathbb N$, with in fact $ |\tWang|_g\big|_{\partial_1 M_i} $ identically zero. But note that the above definition allows for degenerate black holes, such as extreme Kottler black holes with higher-genus topology, which contain asymptotically cylindrical ends along which both  $V$ and $|dV|_g$ tend to zero when receding to infinity along the end, with both the area of the crossections of the cylindrical end and $ |\tWang|_g\big|_{\partial_1 M_i} $ approaching finite non-zero  limits.

We have the following
extension of Theorem~\ref{T11VI19.1}:

\begin{Theorem}
 \label{T11VI19.2}
Consider a  strictly static asymptotically locally hyperbolic solution $(M,g,V)$ of the static Einstein equations
with
at least one mild end. Then the total mass is negative or zero, vanishing if and only if $(g,V)$ is given by \eqref{2VII19.21}.
\end{Theorem}

\proof
Applying the divergence identity on $M_i$ and passing with $i$ to infinity one obtains that the sum of the masses of the ALH ends is non-positive. If the mass vanishes we obtain that $g$ is Einstein, and the result follows  from \cite[Theorem~2, case (II,A)]{Tashiro}, compare the discussion after Theorem~\ref{T11VI19.1} above.
%
\hfill
$\Box$
\bigskip

Recall that one of the obstructions, when attempting to prove the positive mass theorem for asymptotically locally hyperbolic manifolds using spinorial methods \emph{\`a la Witten}, is that of existence of nontrivial spinor fields which asymptote to Killing spinors of the asymptotic background near the conformal boundary at infinity. Such spinor fields will be called \emph{asymptotic Killing spinors}. We shall say that an asymptotically locally hyperbolic spin manifold $(M,g)$ has a \emph{compatible spin structure} if  all  components of the conformal boundary at infinity admit non-trivial asymptotic Killing spinors.

The BK cuspidal metrics with a flat $h$ provide  examples of manifolds with compatible spin structure. Examples which do not have a compatible spin structure are the Kottler black holes with higher genus topology, or the Horowitz-Myers metrics. (Indeed, if they admitted a compatible spin structure, they would all have positive mass, but some of them don't.)

Theorem~\ref{T11VI19.2} leads to the following uniqueness theorem for the BK cuspidal metrics, seemingly unnoticed in the literature so far. To avoid ambiguities: we assume here and below that $M$ has no boundary.

\begin{Theorem}
 \label{T11VI19.3}
Let  $(M,g,V)$ be  a strictly static  solution of the vacuum Einstein equations
which is the union of a finite number of mild
 ends (at least one),   a finite number of ALH ends (at least one), and of a compact set. If $(M,g)$ carries a  compatible spin structure,
 then
$(M,g,V)$ is the cuspidal BK metric \eqref{2VII19.21}.
\end{Theorem}

\proof
Choose any of the asymptotically locally hyperbolic (ALH) ends of $(M,g)$.
One can run the generalisation of Witten's proof of the positive energy theorem as in \cite{Wang,ChHerzlich},  using spinor fields which asymptote to a non-trivial Killing spinor in the chosen end and to zero on all other ends (if any), to conclude that the mass of each ALH end is positive or vanishes. The result follows by Theorem~\ref{T11VI19.2}.
\hfill $\Box$

\begin{remark}
  \label{R4IX19.1}
\emph{
An  example,  not covered by the analysis so far, of a static but \emph{not strictly static} ALH metric with zero mass is the  ``hyperbolic Einstein-Rosen bridge'',
\begin{equation}\label{4IX19.1}
  g= dr^2 + \cosh^2(r) h
  \,,
  \quad
   V = \sinh(r)
   \,,
\end{equation}
with $r\in \R$, where $h$ is a negatively curved Einstein metric  on a compact manifold.
In this case $(M,g,V)$ has two ALH ends. We are not aware of a positive-energy theorem which would hold for this topology, and which could lead to a uniqueness theorem for this metric
using the methods here.
(See~\cite{WangPE} for a uniqueness result for the metric~\eqref{4IX19.1} within the class of Einstein metrics.)
}
\qed
\end{remark}

\medskip

A completely different uniqueness theorem for the cuspidal BK metrics, without spin assumptions, has been recently proved by the second  author and H.C.~Jang \cite{GallowayWarpedSplit}.
 The results there are motivated by the fact
 that  the level sets of the coordinate $r$ of \eqref{2VII19.21} have mean curvature $H=n-1$, so that one can cut the manifold along any such set to obtain a conformally compactifiable manifold with boundary satisfying $H=n-1$.
To obtain a result like Theorem \ref{T11VI19.3}, but without spin assumption, we will make use of the following
 slight refinement of part 3  (the toroidal case)  of Theorem~1.1 in \cite{CGNP}.

 \begin{Theorem}
 \label{cgnptorus}
 Let (M,g) be an $n$-dimensional, $4 \le n \le 7$, asymptotically locally hyperbolic manifold
with flat toroidal conformal infinity $(N, \mathring{h})$, such that $M$ is diffeomorphic to
 $[r_0, \infty) \times N$.  Suppose that:

\begin{enumerate}
\item The boundary $N_0 = \{r_0\} \times N$ has mean curvature $H  \le n-1$, where $H$ is the divergence  $D_i N^i$ of the unit normal $N^i$ pointing into $M$.
\item  The scalar curvature $R$ of $M$ satisfies $R \ge -n(n-1)$.
\end{enumerate}
Then $(M,g)$ has nonnegative mass, $m \ge 0$.
 \end{Theorem}

\noindent
{\it Comment on the proof.}  The only difference in this version is that the condition $H < n-1$ in \cite[Theorem~1.1, part 3]{CGNP} has been replaced by the condition $H \le n-1$.  To explain this weakening, we indicate briefly how the proof goes.  Suppose by contradiction the mass is negative.  Then, as in the proof of  \cite[Theorem~1.1, part 3]{CGNP} there exists a compact   hypersurface $N_1$ out near infinity, cobordant to $N_0$, with mean curvature $H_1 > n-1$.  Then, with respect to the initial data set  $(M, g, K = -g)$, $N_0$ has null expansion $\theta_ 0\le 0$  (and $<0$ if $H < n-1$), and $N_1$ has null expansion $\theta_1 > 0$, both with respect to the null normal fields pointing towards the ALH end. In  the case  $\theta_0 < 0$, the basic existence result for marginally outer trapped surfaces (MOTS) (see e.g. \cite[Theorem 3.3]{AndEM})  guarantees the existence of an outermost MOTS in the region between $N_0$ and $N_1$. However, as was carefully shown in Theorem~5.1 in \cite{AnderssonMetzger2}, the assumption
$\theta_1 < 0$ can be weakened to $\theta_1 \le 0$. The only difference is that the outermost MOTS $\Sigma$,
whose existence is guaranteed by this theorem, may have some components in common with
$N_0$.  Now, as discussed in the proof of \cite[Theorem~1.1, part 3]{CGNP},  $\Sigma$ (or some component of
$\Sigma$) cannot carry a metric of positive scalar curvature.  But then Theorem~3.1 in \cite{motsv4}
  implies that
$\Sigma$ cannot be outermost.  Hence the mass must be nonnegative.

\smallskip

We further remark, as was similarly noted in \cite{CGNP}, the condition that
$(N, \mathring{h})$ is a flat torus can be replaced by the somewhat more general condition that $(N, \mathring{h})$  is a compact flat manifold,  provided the product assumption in Therorem \ref{cgnptorus} extends to the conformal boundary. This follows from a covering space argument, using the fact that any compact flat manifold is finitely covered by a flat torus.

\smallskip
Using Theorem \ref{cgnptorus}, one can now argue in a manner similar to the proof of Theorem \ref{T11VI19.3} to obtain the following:

\begin{Theorem}
 \label{T25VI19.1}
Let  $(M,g,V)$ be  a  strictly static  solution of the vacuum Einstein equations diffeomorphic to $\T^{n-1}\times \R$, $4\le n \le 7$, where $ \T^{n-1}$ is a torus,
with one mild end and one asymptotically locally hyperbolic end.
If there exists $r_0\in\R$ such that $\T^{n-1}\times \{r_0\}$ has mean curvature satisfying $H\le n-1$ with respect to the normal pointing towards   the ALH end, then
$(M,g,V)$ is the
BK cuspidal solution $\left((0, \infty) \times \T^{n-1}, r^{-2} dr^2 + r^2 h\right)$.
\end{Theorem}

The general BK rigidity result obtained in recent work of L.-H. Huang and H.C. Jang \cite{LanHyun} (see
Remark  \ref{R2IV19.1} below)  involves ALH manifolds with compact boundary, and without ``internal" cuspidal ends.   In a similar vein, we consider below a uniqueness result for the BK cuspidal spaces in the context of static vacuum ALH manifolds with boundary.  This result makes use of certain properties of constant mean curvature (CMC) hypersurfaces in Riemannian manifolds, which we now describe; cf., e.g., \cite{MeeksAndCo}.

Let $\Sigma$ be a two-sided compact hypersurface in a Riemannian manifold $(M,g)$ of dimension $n$.  Hence,
$\Sigma$ admits a smooth unit normal field $N$.   Consider a normal variation $t \to \Sigma_t$  of
$\Sigma = \Sigma_0$, i.e. a variation with variation vector field
$V = \frac{\partial}{\partial t}|_{t=0} = \phi N$.  $\phi \in C^{\infty}(\Sigma)$.
Let $\mathcal{B}(t) = \mathcal{A}(t) - (n-1)\mathcal{V}(t)$, where $\mathcal{A}(t)$ is the area of
$\Sigma_t$ and $\mathcal{V}(t)$ is the (signed) volume of the region bounded by $\Sigma_t$ and $\Sigma$.  Then a computation shows that $\Sigma$ has mean curvature $H = n-1$ if and only if $\mathcal{B}'(0) = 0$ for all normal variations $t \to \Sigma_t$.   We say that $\Sigma$ is a {\it stable} CMC hypersurface, with mean curvature $H = n-1$, provided $\mathcal{B}''(0) \ge 0$ for all normal variations $t \to \Sigma_t$.
Consider the operator $L: C^{\infty}(\Sigma) \to C^{\infty}(\Sigma)$, defined by
\begin{equation}
L(\phi) = - \Delta \phi + \frac12(R_{\Sigma} - R - |A|^2 - H^2)\,\phi
\end{equation}
where, as before, $A$ is the second fundamental form of $\Sigma$.
It is a well known fact that $\Sigma$ is stable if and only if the principal eigenvalue of $L$ is nonnegative,
$\lambda_1(L) \ge 0$.  We will take this analytic characterization as our definition of stability.
Using this characterization, the following was proved in \cite{AnderssonGallowayCai}.

\begin{Lemma}\label{infrigid}
Let $(M,g)$ be an $n$-dimensional Riemannian manifold
with scalar curvature $S$ satisfying,
$S \ge -n(n+1)$.
Let $\Sigma$ be a compact $2$-sided  stable CMC hypersurface in $M$, with mean curvature $H = n-1$.  Suppose $\Sigma$ does not carry a metric of positive scalar curvature.   Then the following holds.
\begin{enumerate}
\item[(i)] $\Sigma$ is umbilic, in fact $A =h$,
where $h$ is the induced metric on $\Sigma$.
\item[(ii)] $\Sigma$ is Ricci flat and $S = -n(n+1)$ along $\Sigma$.
\end{enumerate}
\end{Lemma}

We now consider the following uniqueness result for the cuspidal BK space.

\begin{Theorem}\label{rigid}
Let $(M,g)$ be a strictly static ALH manifold with compact boundary $\Sigma$, and with  static potential $V$, such that $(M,g,V)$ satisfies the static vacuum Einstein equations.
Suppose that:
\begin{enumerate}
\item $\Sigma$ is a stable CMC hypersurface with mean curvature $H_{\Sigma} = n-1$ (with respect to the inward pointing unit normal $N$).
\item $\Sigma$ does not carry a metric of positive scalar curvature.
\item Conditions hold which imply that $(M,g)$ has nonnegative mass, $m \ge 0$ (compare
the positivity results in \cite{ChHerzlich} for manifolds with compatible spin structure, or
Theorem~\ref{cgnptorus} above.)
\end{enumerate}
Then $m = 0$ and $(M,g)$ is Einstein, $R_{ij} = -(n-1)g_{ij}$.  Furthermore, if $\Sigma$ is a regular level set of the static potential, $\Sigma = \{V = V_0\}$, then $(M,g)$ is isometric to the (truncated) cuspidal BK space $([r_0, \infty) \times \Sigma, r^{-2} dr^2 + r^2 h)$.
\end{Theorem}

\proof  Using the Gauss equation \eqref{scalarconstraint}, together with Lemma \ref{infrigid}, one easily computes,
\begin{equation}
R_{ij}N^iN^j = -(n-1) \,.
\end{equation}
Furthermore, by the Codazzi equation and part (i) of Lemma \ref{infrigid}, one has
\begin{equation}
R_{ij}X^iN^j = 0 \quad  \text{for all  $X$ tangent to $\Sigma$} \,.
\end{equation}

Now, applying \eqref{integration}, we obtain
\begin{align}
\int_M V|W|^2 d\mu &= \int_{\partial M_{\infty}} W_{ij}D^iVN^j d\sigma + \int_{\Sigma} W_{ij}D^iVN^j d\sigma_{\Sigma}  \nonumber \\
&= -8(n-2)m \pi + \int_{\Sigma} W_{ij}D^iVN^j d\sigma_{\Sigma}  \,;
\end{align}
recall that $W_{ij}  = R_{ij} + (n-1)g_{ij}$.

Along $\Sigma$ we can write $D V$ as,
\begin{equation}
D V = \lambda N + X
\end{equation}
where $X$ is tangent to $\Sigma$.  Hence, along $\Sigma$, we have
\begin{align*}
W_{ij}D^iVN^j &= \lambda W_{ij}N^iN^j  + W_{ij}X^iN^j  = \lambda W_{ij}N^iN^j   \\
&= \lambda (R_{ij}N^iN^j + (n-1)g_{ij}N^iN^j) \\
&=  \lambda ( -(n-1) + (n-1) )= 0 \,.
\end{align*}
Thus,
\begin{equation}\label{mass}
\int_M V|W|^2 d\mu = -8(n-2)m \pi .
\end{equation}
and hence $m \le 0$.  Assumption 3  in Theorem  \ref{rigid} then implies $m = 0$, and hence by \eqref{mass},  $R_{ij} = -(n-1) g_{ij}$.

We now assume $\Sigma$ is the level set, $\Sigma = \{V = V_0\}$, and show that $(M,g)$ is isometric to the cuspidal BK space, as in the statement of the theorem.  We have,
\begin{equation}\label{hess}
D_iD_j V = V (R_{ij} + n g_{ij}) = Vg_{ij}  \,.
\end{equation}
Hence, in view of Proposition 4.2 in \cite{GallowayWarpedSplit}, it suffices to show that, along $\Sigma$, $|D V| = V_0$.

Along $\Sigma$, we have $D V = \lambda N$, where $\lambda = \pm |D V|$.  Let $X$ be any unit tangent vector to
$\Sigma$.  From \eqref{hess}, $D_iD_jVX^iX^j = V_0$.  On the other hand, using the definition of the Hessian,
\begin{align*}
D_iD_jVX^iX^j   &= g_{ij}X^kD_kD^iVX^j   = g_{ij}X^kD_k(\lambda N^i)X^j  \\
&=\lambda g_{ij}X^kD_k N^iX^j   + (X^kD_k\lambda) g_{ij}N^iX^j \\
&= \lambda A_{ij}X^iX^j = \lambda h_{ij}X^iX^j = \lambda\,,
\end{align*}
where in the last line we have used part (i) of Lemma \ref{infrigid}.  Thus $\lambda = V_0 > 0$, and hence $|D V| = V_0$ along $\Sigma$.\qed

\begin{Remark}
  \label{R2IV19.1}\emph{ Theorem 4.1 in \cite{GallowayWarpedSplit} shows that under a strengthening of the stability assumption, the assumption in Theorem~\ref{rigid} that the boundary is a level set of the static potential, used to conclude that $(M,g)$ is isometric to the cuspidal BK space, can be removed.  This strengthened assumption (``locally weakly outermost") is used in forthcoming work of Lan-Hsuan Huang and Hyun Chul Jang \cite{LanHyun}, in which they establish the uniqueness of the cuspidal BK space
    in a more general (not necessarily static) ALH setting with boundary, assuming the mass vanishes.  We thank them for communications regarding their work.}

\end{Remark}


\section{Lower bound for  entropy (area),   spherical conformal infinity}

We continue to assume that $\Lambda <0$ but now we consider solutions with a horizon $H$. If the conformal boundary at infinity is a sphere then by the Riemannian asymptotically hyperbolic positive mass theorem \cite{chrusciel_hyperbolic_2019} we have $m\geq0$ so
\begin{eqnarray}
 \lefteqn{\sum_{H_p}\kappa_p\int_{H_p}(\redRH  +(n-2)(n-1))\,\mathrm{d}\sigmaH }
 \nn
\\
 & = & 2\int_MV|\tWang|_g^2\,\mathrm{d}\mu_g
 +   \nored{{16 (n-2) m \pi }}
 \geq 0
 \,,
 \label{7VII19.2}
\end{eqnarray}
and the inequality is saturated if and only if $m=0$ and $\tWang=0$. This last condition is equivalent to, since $\varepsilon = -1$,
\begin{equation}
 \redR _{ij} = -(n-1)g_{ij}
 \,,
\end{equation}
so $(M, g)$ is an Einstein manifold (the Ricci tensor is proportional to the metric).

If the manifold $M$ is $3$-dimensional, $n=3$, the identity \eqref{7VII19.2} implies that
\begin{equation}
 \sum_{H_p}\kappa_p\int_{H_p}(\redRH  + 2)\,\mathrm{d}\sigmaH  \geq 0
 \,,
\end{equation}
that is to say
\begin{equation}
 2\sum_{H_p}\kappa_p A_p \geq - \sum_{H_p}\kappa_p\int_{H_p}\redRH \,\mathrm{d}\sigmaH
 \,,
\end{equation}
where $A_p$ is the area of the connected component $H_p$ of the boundary $H$.

Assuming that the $H_p$'s are  all closed and orientable, the Gauss-Bonnet theorem,
\begin{equation}
\int_{H_p}\redRH \,\mathrm{d}\sigmaH  = 
8 \pi (1-g_p)
 \,,
\end{equation}
where $g_p$ is the genus of $H_p$, gives the inequality
\begin{equation}
 \sum_{H_p}\kappa_pA_p \geq 4\pi\sum_{H_p}\kappa_p(g_p-1)
 \,.
\end{equation}
In order to express this inequality in terms of the cosmological constant $\Lambda$ we have to use the fact that according to \eqref{lambdanorm} for $n=3$ we set $\Lambda = -3$, so in this case we have
\begin{equation}
 \sum_{H_p}\kappa_pA_p \geq \frac{12\pi}{|\Lambda|}\sum_{H_p}\kappa_p(g_p-1)
 \,.
 \label{ineqarea}
\end{equation}
Note that the cosmological constant $\Lambda$ has the dimension of  {inverse length squared} so this is the only way to reintroduce it while preserving the homogeneity of the dimensions.

 In the case where $H$ is connected this reads
\begin{equation}
 A_H \geq \frac{12\pi (g_H - 1)}{|\Lambda|}
 \,,
 \label{AreaBoundNeg}
\end{equation}
with $A_H$ the area of $H$ and $g_H$ its genus. This can be compared to a weaker inequality of  Gibbons \cite[Equation~(45)]{gibbons_comments_1999},  where time symmetry but no staticity is  assumed
  (see also Woolgar \cite{woolgar_bounded_1999})
\begin{equation}
 A_H \geq \frac{4\pi(g_H-1)}{|\Lambda|}
 \,.
\end{equation}

\section{An upper bound for the free energy}
 \label{7VII19}


 Let $k\in \R$. In \cite{CGW} one defines
 \begin{equation}\label{7VII19.5}
  F_k = E- k T S
  \,,
 \end{equation}
 where $E $ is the total mass,  $T$ is the Hawking temperature of a Killing horizon, and $S$ its entropy~\cite{GibbonsHawkingCEH}
 \begin{equation}\label{7VII19.1}
 E = m
 \,,
  \quad
   T = \frac{\kappa}{2\pi}
    \,,
    \quad
    S = \frac{A}{4}
     \,.
 \end{equation}
 The functional  $F_k$ equals the total mass when $k=0$  and the ``free energy"  when $k=1$.

From \eqref{7VII19.2} we have
\begin{equation}
 \sum_{H_p}\kappa_p\int_{H_p}(R(h) +(n-2)(n-1))\,\mathrm{d}\sigmaH  =
 2\int_MV|F|_g^2\,\mathrm{d}\mu_g
  +\nored{16 (n-2) m \pi } 
 \,,
 \label{7VII19.2again}
\end{equation}
which
in dimension $n=3$ and for a single Killing horizon $H$ reads
\begin{equation}
 \label{7VII19.3}
  2 \pi T
  \left(
   \int_{H}R(h) \,\mathrm{d}\sigmaH
   + 2 A
    \right)
 -   16 \pi m
 \geq 0
 \,,
\end{equation}
If the Killing horizon is a torus or an orientable manifold of higher genus $g_H$ (not to be confused with the metric $g_H$ on the horizon...) we obtain
\begin{equation}
 \label{7VII19.4}
  E
  -
    T S \leq 8 \pi  \kappa  (1-g_H )
  \leq 0
 \,.
\end{equation}
 Using the notation of \eqref{7VII19.5}, we conclude that the free energy $F\equiv F_1$ of static solutions containing a connected horizon of higher genus is negative:
 \begin{equation}\label{7VII19.6}
  F_{1} \le 0
  \,.
 \end{equation}
This is stronger than the inequality $F_{2/5}\le 0$ of~\cite{CGW},  with a different proof. 

\section{Positive cosmological constant}

In this section we consider a positive cosmological constant so $\Lambda>0$ and $\varepsilon =1$ in \eqref{lambdanorm}. In this case there is no conformal boundary at infinity, and the models $(M,g,V)$ of interest are compact manifolds with a boundary $H$ on which  $V$ vanishes. Thus \eqref{integration2} becomes
\begin{equation}
 \int_MV|\tWang|_g^2\,\mathrm{d}\mu_g
 = \sum_{H_p}\frac{\kappa_p}{2}\int_{H_p}(\redRH  - (n-1)(n-2))\,\mathrm{d}\sigmaH
 \geq 0
 \,,
\label{integrationpos}
\end{equation}
and the inequality is saturated with a non-trivial $V$ if and only if $\tWang=0$ on $M$, that is to say if and only if
\begin{equation}
 \redR _{ij} = (n-1) g_{ij}
 \,.
\end{equation}

If $M$ is $3$-dimensional, $n=3$, from \eqref{integrationpos} we obtain
\begin{equation}
 \sum_{H_p}\kappa_pA_p \leq \frac{1}{2}\sum_{H_p}\kappa_p\int_{H_p}\redRH \,\mathrm{d}\sigmaH
 \,.
\end{equation}
Assuming the $ H_p $'s are all closed and orientable and applying the Gauss-Bonnet theorem
we are led to
\begin{equation}
 \sum_{H_p}\kappa_pA_p \leq 4\pi\sum_{H_p}\kappa_p(1-g_p)
 \,.
\end{equation}
Reintroducing the cosmological constant ($\Lambda = +3$ so far) yields
\begin{equation}
 \sum_{H_p}\kappa_pA_p \leq \frac{12\pi}{\Lambda}\sum_{H_p}\kappa_p(1-g_p)
 \,.
 \label{IneqPos}
\end{equation}
If $H$ is connected then this inequality becomes
\begin{equation}
 A_H \leq \frac{12\pi}{\Lambda}(1-g_H)
 \,,
\end{equation}
and since $A_H > 0$ and $g_H \in \mathbb{N}$ we must have
\begin{equation}
 g_H = 0
 \,,
\end{equation}
so the genus of $H$ is zero, which means that $H$ is homeomorphic to a two-sphere, and we get
\begin{equation}
 A_H \leq \frac{12\pi}{\Lambda}
 \,.
\end{equation}
If one sets $\Lambda = 3$ through a constant conformal rescaling then the result becomes
\begin{equation}
 A_H \leq 4\pi
 \,.
\end{equation}
The above\footnote{Once this paper was written we realised that Theorem~\ref{T30VIII19.1}, as well as the proof presented here, can already be found in \cite{Shen}.}
 gives a simple proof of a result  already  obtained by Boucher-Gibbons-Horowitz \cite[Equation~(3.1)]{boucher_uniqueness_1984} (compare Ambrozio \cite[Theorem~2]{ambrozio_static_2015} and Borghini-Mazzieri \cite[Corollary~2.6]{BorghiniMazzieri1}):

\begin{Theorem}
\label{T30VIII19.1}
Let $(M, g, V)$ be a $3$-dimensional compact static solution to the vacuum Einstein equation with positive cosmological constant $\Lambda$. If the boundary $H$ of $M$ is connected, closed and orientable then $H$ is homeomorphic to a two-sphere and its area $A_H$ satisfies
\begin{equation}
 A_H \leq \frac{12\pi}{\Lambda}
 \,,
 \label{AreaBoundPos}
\end{equation}
with equality holding only in de Sitter space.
\end{Theorem}

We note that the equality case is handled as before by an analysis of Obata's equation, and that the condition of orientability can be removed by passing to a finite covering of $M$, in which the areas of each horizon will be larger than or equal to the original ones.

\appendix
\

\section{The mass of the Birmingham-Kottler metrics}

Consider a Birmingham-Kottler (BK) metric~\cite{Birmingham,Kottler} in space-time dimension $d\equiv n+1$:
\begin{equation}\label{9VII19.1}
  g = - V^2 dt^2 + \frac{dr^2 }{V^2} + r^2\underbrace{ h_{AB} dx^A dx^B}_{=: h_k}
  \,,
  \quad
  V^2 = \frac{r^2}{\ell^2} + k - \frac{2 \mu }{r^{n-2}}
   \,,
\end{equation}
with $\mu \in \R$,
where $\ell$ is related to the cosmological constant $\Lambda$ as
$$
 \Lambda = -\frac{n(n-1)}{2 \ell^2}
  \,,
$$
and where $h_k$ is an Einstein metric with scalar curvature $(n-1)(n-2) k$, with $k\in \{\pm 1, 0\}$. When $k\ne 0$ the coordinates above are uniquely defined except when $h_k$ is the unit round metric on a sphere $S^{n-1}$, in which case the coordinates are defined up to a conformal transformation of the sphere.
On the other hand the case $k=0$ allows for the rescaling
\begin{equation}\label{15VII19.1}
 (r,t) \mapsto (\bar r = \lambda r, \bar t =  \lambda ^{-1}t)
 \,,
\end{equation}
where $\lambda$ is a positive constant, in which case \eqref{9VII19.1} becomes
\begin{eqnarray}\label{9VII19.4}
  g
    & = &
    - \bar V^2 d\bar t^2 + \frac{d\bar r^2 }{\bar V^2} + \bar r^2\underbrace{ \bar h_{AB} dx^A dx^B}_{=: \bar h_k}
  \,,
\\
  \bar V^2
   & = &
    \frac{\bar r^2}{\ell^2}  - \frac{2\bar \mu}{\bar r^{n-2}}
   \,,
\end{eqnarray}
with
\begin{equation}\label{9VII19.2}
  \bar h_k = \lambda^{-2} h_k
  \,,
   \quad
  \bar \mu = \lambda^n \mu
  \,,
   \quad
   \mbox{and note that } \
  \bar V = \lambda V
   \,.
\end{equation}
If we take the point of view that only the conformal class of $h_k$ matters as far as the asymptotic data are concerned, we conclude that, when $k=0$, the only information carried by the number  $\mu$ is its vanishing or its sign.

Now, we see from \eqref{9VII19.2} that
\begin{equation}\label{9VII19.3}
 \bar V \times \bar \mu \times   \sqrt{\det  \bar  h_k} =
 V \times \mu \times   \sqrt{\det  h_k}
\end{equation}
so that
 the product
\begin{equation}\label{9VII19.6}
 V \times \mu \times   \sqrt{\det  h_k}
  \,,
\end{equation}
which has the same scaling as the integrand of the formula \eqref{8VII19.1} defining the Hamiltonian energy,
is invariant under \eqref{15VII19.1}.

In conclusion, 1) $\mu$ by itself is \emph{not} the Hamiltonian mass, and 2)  the inclusion of $V$ and of the area factor associated with the metric $h_k$ in the integrand are essential for an invariant definition of mass.

\subsection{The space Ricci tensor for Birmingham-Kottler metrics}
 \label{ss22VII19.1}

We wish to calculate the boundary integral at infinity which arises when integrating the divergence identity \eqref{31VIII19.1} for a BK metric. For this we need the space-part of its Ricci tensor.
The simplest way to calculate this tensor is to use the static KID equation \eqref{Einsteingnorm} written backwards
\begin{equation}\label{18VII19.1}
  R_{ij }= V^{-1}D_i D_j V  +   \frac{2\Lambda}{n-1}g_{ij}
   \,.
\end{equation}
One readily finds, in the scaling where $2\Lambda = - n(n-1)$, in an ON frame with $\theta^{\hat 1} = V ^{-1}dr$,
and where $\theta^{\hat A} $ is orthonormal frame for  the metric $r^2 h_k \equiv r^2 h_{AB} dx^A dx^B$,
\begin{eqnarray}
  R_{\hat 1 \hat 1}
   & = &
    V^{-1}D_{ \hat 1} D_{\hat 1} V - n g_{\hat 1 \hat 1}
    \nn
\\
  & = &
  1-n- \frac{(n-2) (n-1) m }{ r^{n}}
   \,,
\\
  R_{\hat A \hat B}
   &= &
    V^{-1}D_{ \hat A} D_{\hat B} V - n g_{\hat A \hat B}
    \nn
\\
 & = &
  \left(
  1-n+ \frac{(n-2)  \mu }{ r^{n}}
   \right)
   \delta_{\hat A \hat B}
  \,,
  \label{18VII19.2}
\end{eqnarray}
with the remaining components of the Ricci tensor being zero by symmetry considerations. These formulae readily lead to (compare
\cite{HerzlichRicciMass})
%
\begin{eqnarray}
   \lefteqn{-
   \lim_{R\rightarrow\infty}\int_{r=R}  \nabla ^j V
    ( R^i{}_j - \frac R n \delta^i_j)
    d\sigma_i}
    &&
    \nn
   \\
     & & \phantom{xxxxxx}= {(n-1)(n-2)}A_\infty \mu
     \,,
           \label{11VII19.1x}
\end{eqnarray}
where $A_\infty$ is the area of the boundary at infinity in the metric $h_k$.

\bigskip
\bibliographystyle{unsrt}

\bibliography{ChruscielPotaux-minimal}
\end{document}